# Damage due to salt crystallization in porous media

*Noushine Shahidzadeh Bonn*[*,1,2], *François Bertrand*[2], *Daniel Bonn*[1,3],

1-Van der Waals-Zeeman Instituut (WZI), Universiteit van Amsterdam, Valckenierstraat 65, 1018 XE Amsterdam, The Netherlands

2-Université Paris-Est, UR Navier, LMSGC, 2 allée Kepler, 77420 Champs-sur-Marne, France

3-Laboratoire de Physique Statistique de l'ENS, 24 rue Lhomond 75231 Paris cedex 05, France.

**Abstract**

We investigate salt crystallization in porous media that can lead to their disintegration. For sodium sulfate we show for the first time experimentally that when anhydrous crystals are wetted with water, there is very rapid growth of the hydrated form of sulfate in clusters that nucleate on anhydrous salt micro crystals. The molar volume of the hydrated crystals being four times bigger, the growth of these clusters can generate stresses in excess of the tensile strength of the stone and lead therefore to damage.

Damage due to salt crystallization in porous media is of both fundamental and practical interest. From a practical point of view, many ancient buildings suffer from damage to the stones due to salt crystallization; the most well-known example probably being the Petra monument (Jordania) that has been partially destroyed by salt damage [1]. In addition, climate change is already affecting the conservation of world cultural heritage but also of modern constructions [2]. Since virtually all building constructions (ancient and modern) are built with porous materials that contain salt, any increases in soil moisture or extreme weather conditions (rain penetration) might result in greater salt mobilization. Salts can be naturally present in the stones, get trapped inside the porous material for instance by imbibition with salt-containing precipitation (acid rain), or be present in the mortars used for construction. Drying of wet stones will cause salt crystallization that is believed to be responsible for the deterioration of building materials.

From a fundamental point of view, salt deterioration of porous media has attracted a lot of attention over the past few decades, but remains incompletely understood [3-10]. It is anticipated that a better understanding of the mechanisms involved can significantly improve the methods used for preventing damage. Different theoretical explanations have been proposed, the most popular being the crystallization pressure of supersaturated salt solutions. These form when a salt-containing stone is wet (e.g., by rain) and subsequently dries out. The supersaturation can account for an excess pressure exerted by the salt crystals against the pore walls [5-8]. However, experimentally it is still not clear, why under identical experimental conditions, certain salts such as sodium sulfate cause much more damage than other high-solubility salts such as sodium chloride, although sodium chloride has the higher crystallization pressure [5,8]. Consequently, a detailed understanding of how crystal growth within the porous media leads to damage still remains elusive.

In this Letter, we investigate the crystallization of sodium sulfate during wetting and drying cycles in a porous medium and compare it with the case of sodium chloride. Using optical microscopy and Magnetic Resonance Imaging (MRI) in the stones, we show that for the sulfate, a rapid formation of hydrated salts during wetting/drying cycles is the major cause of the mechanical damage observed in the sandstone. Sodium sulfate has two stable crystal phases at room temperature: the anhydrous salt ($Na_2SO_4$, thenardite) and the decahydrated form ($Na_2SO_4 \cdot 10H_2O$, mirabilite). Sodium chloride has only a stable anhydrous phase. For the sulfate, we provide unambiguous experimental evidence that when anhydrous crystals are wetted with water, the growth of hydrated crystals happens in a manner that is very different from the formation of the same crystals from a supersaturated solution. This happens because the anhydrous crystals only partly dissolve and subsequently act as nucleation sites for the formation of a large amount of hydrated crystals that form in clusters. The molar volume of the hydrated crystals being four times bigger, we show that the growth of these clusters can generate stresses in excess of the tensile strength of the stone and lead therefore to damage.

We perform experiments on both macroscopic and microscopic scales. The macroscopic measurement consists of a simultaneous study of the global drying kinetics by weight measurements and the local water saturation using MRI measurements; both are done simultaneously during the evaporation of water from porous media containing salts. Prague (Mšené) sandstones of porosity $\phi \sim 21\%$ and average pore diameter $d_p=30$ μm were used as the porous medium. Samples (2.5x2.5x4 cm) were cut from the same block, were saturated by imbibition with the different salt solutions, and left to evaporate. Since sodium sulfate has two forms, its solubility in water depends on the crystalline form considered. Then, saturated (16.5 wt%, saturation S=1 (w.r.t. mirabilite) and different supersaturated (up to 20.5 %wt S=1.25) solutions were prepared at 21ºC by dissolving thenardite (SigmaAldrich grade) in ultrapure water [11]. Saturated sodium chloride solutions (26.4 wt%) were also prepared. After the first drying cycle, part of the salt will crystallize within the porous medium (so-called subflorescence) and part at the surface of the porous medium (efflorescence). The efflorescence is removed before starting a second cycle of imbibition with pure water and drying. All experiments were done under at T=21°C, Relative Humidity (RH=48±5%). In both cycles, the weight of the stone is followed in time during drying on an automated balance with a precision of (±0.001 g).

For the first cycle (C1), imbibition of the stone was done with either saturated (S~1) or supersaturated (1< S <1.25) salt solutions. The first important observation is that there is no damage observed at this stage, neither for the sulfate, nor for the chloride. For the second cycle (C2), the salt-containing stones were imbibed with pure water and left to dry under the same environmental conditions. After this second cycle of drying a rather impressive disintegration of the stone can be observed in the case of sodium sulfate (Fig 1). The damage for the sulfate can be up to 10-12% of the initial mass of the stone. No measurable damage is observed for the case of sodium chloride although a larger amount of salt was present in the stone. The damage was assessed by brushing the surface, and subsequently washing and drying the stone, to determine its mass relative to the initial one. The surprising observation is that damage only occurs in the second cycle, implying that is not simply due to the crystallization of salt in the porous medium, which also happens during C1.

The question is now what the mechanism of the damaging is. To get an idea of the order of magnitude of the tensile strength, we perform a simple three-point flexion experiment [13], in which we determine the maximum force the sample can withstand without breaking. For a rectangular sample under a load in a three-point flexion setup the stress is: $\sigma = \frac{3FL}{2bd^2}$, where $F$ is the load (force), $L$ is the length of the support span, $b$ the width and $d$ the thickness of the rectangular sample. We find the breaking force to be 3.4 N; putting in the numbers for the size, we find that the tensile strength is ~0.9 MPa, in good agreement with previous measurements [14]. Thus, the salt crystallization has to generate internal stresses in excess of this value if it is to break the stone.

MRI was used to follow the dynamics of drying within the stone for each cycle [12]. Water density profiles were obtained along the $z$-axis, with a reproducibility of better than 5%. Fig.2 shows saturation profiles for the two different cycles in the case of sodium sulfate. Just after imbibtion with pure water to start the second cycle, there is approximately 5% less water in the upper part of the porous medium compared to the first cycle, exactly the volume of salt that is present as subflorescence. Therefore the salt is mainly located where the damage occurs (Fig.1). From Fig.2, the strange observation is that e.g., after 5 hours, the water loss appears to be faster during cycle 2 although the drying happens under the exact same environmental conditions and a virtually equal amount of water has been added. A related observation is that there is a serious discrepancy between the rate of evaporation obtained from the MRI profiles and the weight measurements (Fig.2(b)); both measurements are done simultaneously on the same sample. These problems are resolved by realizing that the water does not disappear from the stone, but goes into hydrated crystals that are not visible on the MRI, but do of course contribute to the weight. Indeed, an MRI measurement of hydrated crystals in a beaker shows that the water in those crystals does not contribute to the measured proton density, because their NMR relaxation time becomes very short.

The difference between the MRI and weight measurements consequently allows to calculate the amount of hydrated crystals (mirabilite) formed during drying [15]. In Table 1, we compare the quantity of mirabilite formed in cycle 2 after a first cycle in which two similar stones were imbibed using two different salt solutions; we discuss here the two representative cases S=1.04 and S=1.12 from a total of 20 experiments.

|  | Salt solutions used in C1 ($S=C/C_0$) | Salt remaining in sandstones at the end of C1 (%) | Damage at the end of C2 (%) | Mirabilite formed during C2 after 5h (%) |
| --- | --- | --- | --- | --- |
| Stone 1 | 1.04 | 45 | 2.8 | 34 |
| Stone 2 | 1.12 | 50 | 8.3 | 85 |

Table 1. Results from two representative experiments.

The results suggest that the quantity of mirabilite (hydrated crystal) formed in cycle 2 is directly responsible for the damage, and increases strongly with increasing supersaturation of the solution used in the first cycle. This is borne out also in the data of Fig.1(b), where we show that the amount of damage increases linearly with increasing supersaturation in C1. Thus, the formation of the hydrated crystals is responsible for the damage; this has been suspected for a while [3,4], but had not been demonstrated unequivocally so far.

It turns out that the damage is also related to the way the hydrated crystals form. The combined MRI and weight measurements in fact show that a very significant amount of hydrated crystals also form during the first cycle (typically around 70%), but no damage is observed. Under our experimental conditions, the hydrated crystals lose their water at the end of the first drying cycle: they transform into anhydrous crystals. This suggests that the key lies in understanding how the anhydrous crystals formed after the first cycle transform into hydrated crystals when put in contact with water in the second cycle. To study this transformation, we perform optical microscopy experiments. We follow crystallization using optical microscopy again during 2 subsequent impregnation/drying cycles in rectangular (100x800 μm) glass capillaries as model systems for a single pore. Compared to cylindrical capillaries, these represent the irregular pore structure of real porous media (i.e. stones) more realistically due to the presence of thick wetting films in the corner. For the first cycle, capillaries were saturated with different sodium sulfate solutions and after drying we do a second imbibition with pure water followed by drying.

During C1, we directly see hydrated (mirabilite) crystals forming, as is evident from their crystal structure [12]. These subsequently lose their water; however the macroscopically observed form of the crystal remains the same so that it is likely that they transform into small anhydrous micro crystals (thenardite). For C2, these crystals are rewetted with the same volume of water that was initially present. Upon rewetting, we observe that the dissolution of the anhydrate is only partial and in its place there is a region of highly concentrated solution that forms; the maximum solubility of the hydrated crystal is only half that of the anhydrous crystal so that the solution can be highly supersaturated with respect to the formation of hydrated crystals. The small thenardite crystallites that remain present in the solution are observed to act as seeds (nucleation sites) for large amounts of hydrated crystals that start to grow from the remaining thenardite aggregates, giving rise to structures that bear a similarity to a bunch of grapes (Fig.3); these clusters are observed to expand very rapidly. They grow in fact much more rapidly than the mirabilite crystals observed in the first cycle. Since the growth velocity of crystals is directly related to the supersaturation, this suggests that because of their structure the micro crystals of thenardite dissolve very rapidly, leading to high local supersaturations.

From a measurement of the growth speed of the crystals we can in fact infer the supersaturation. The growth speed of a crystal determined from the microscopy images during C1 is about 0.022 μm/s in the fastest growing crystalline direction (Fig.4). However, the growth rate of the hydrated crystals in the second cycle is more than an order of magnitude larger (Fig.4, inset): 0.26 μm/s. Comparing to the detailed measurements of [16] of the growth velocity as a function of supersaturation, this leads to a supersaturation (saturation above $C_{sat}$= 1.4 mole/l) of ≈0.018 mole/l for the first cycle, and ≈0.21 mole/l for the second cycle.

Is this sufficient to break the sandstone? Steiger [8] calculated the crystallization pressure as a function of supersaturation and finds (for small supersaturations, which pertains to our experimental conditions) a crystallization pressure $\Delta P$ = 6.7 MPa per mole/l of supersaturation of mirabilite. Thus for 0.02 mole/l (C1), the crystallization pressure of mirabilite is $\Delta P$ =0.12 MPa, much smaller that the tensile strength of the sandstone, which was almost 1 MPa. This explains why during the first cycle no damage is observed. On the other hand, in C2, a supersaturation of 0.2 mole/l is reached. In this case $\Delta P$ =1.4 MPa, larger than the tensile stress and is therefore sufficient to break the sandstone, as is observed experimentally. These observations explain for the first time why damage is seen during the second cycle and not when mirabilite crystals are formed directly during the first cycle. It also explains the rather dramatic sensitivity of the damage on the saturation used in the first cycle. A saturated solution (S=1) as was shown previously [11] can lead to formation of both hydrated and anhydrate crystals directly and consequently on average to a smaller quantity of micro crystals that act as seeds in the next cycle.

In conclusion, during the rewetting /drying cycles the damage is severe because of the only partial dissolution of anhydrous crystals in regions (pores) that are highly concentrated in salt. The thenardite micro crystals dissolve very rapidly, and in part act as seeds to form large amount of hydrated crystals creating grape-like structures that expand a factor of four in volume. These clusters generate stresses larger than tensile strength of the stone, which leads to the damage. All these effects related to the existence of both hydrated and anhydrous forms, and are consequently absent in salts such as sodium chloride.

Acknowledgements. LPS de l'ENS is UMR8550 and LMSGC is UMR113 of the CNRS. We thank B. Lubelli for the sandstones, and O. Coussy and X. Chateau for helpful discussions.

**REFERENCES**

1. Nicolaescu, A. "Conference Review: Salt Weathering on Buildings and Stone Sculptures", *e-conservation magazine*, 8, 6 (2009)
2. *UNESCO World Heritage Centre publications on climate change* CLT-2008/WS/6 (2008).
3. Goudies, A.S.; Viles, HA. *Salt weathering hazard* (Wiley London, 1997)
4. Rodriguez-Navarro, C.; Doehne, E.; Sebastian, E. *Cement and Concrete Research 30,* 1527 (2000)
5. Flatt, R.J. *J. Cryst. Growth*, *242*, 435 (2002)
6. Tsui, N. , Flatt, R.; Scherer, G.W. *J. Cult. Heritage*, *4*, 109 (2003)
7. Scherer, G.W. *Cement and Concrete Research*, *34*, 1613(2004); Coussy, O, *J. Mech.Phys. Solids*, *54*, 1517 (2006)
8. Steiger, M. *J. Cryst. Growth*, 282, 455 (2005)
9. Rijners, L.A.; Huinink, H.P.; Pel, L.; Kopinga, K. *Phys. Rev. Lett. 94*, 75503 (2005)
10. Espinosa Marzal R.M., Scherer G.W., *Environ Geol* 56, 605 (2008)
11. Shahidzadeh-Bonn, N., Rafai S, Bonn D., Wegdam G, *Langmuir,* 24, 8599 (2008)
12. Bonn, D., Rodts, S., Groenink, M., Rafai, S., Shahidzadeh-Bonn, N., Coussot, P. *Annu. Rev. Fluid Mech*. 40, 209 (2008)
13. Shahidzadeh-Bonn N., Vié P., Chateau X., Roux J-N., Bonn D, *Phys.Rev.Lett*. 95 175501 (2005)
14. This agrees with previous determination of the mechanical properties: tensile strength 0.9-1.6 MPa Pavlík, Z. - Michálek, P. - Pavlíková, M. - Kopecká, I. - Maxová, I. *Construction and Building Materials*, 22, 1736 (2008)
15. It should be noted that although the formation of metastable heptahydrated phase has been reported in cooling experiments in these experiments done at room temperature we only consider the decahydrate phase, as the only stable hydrated phase of sodium sulfate; see [9,10] and Hamilton A, Hall C., Pel L. *J. Phys. D: Appl.Phys.* 41 (2008) for a recent discussion of the heptahydrate.
16. D. Rosenblatt, S.B. Marks, R.L. Pigford*, Ind. Eng. Chem. Fundamen*.23, 143 (1984)

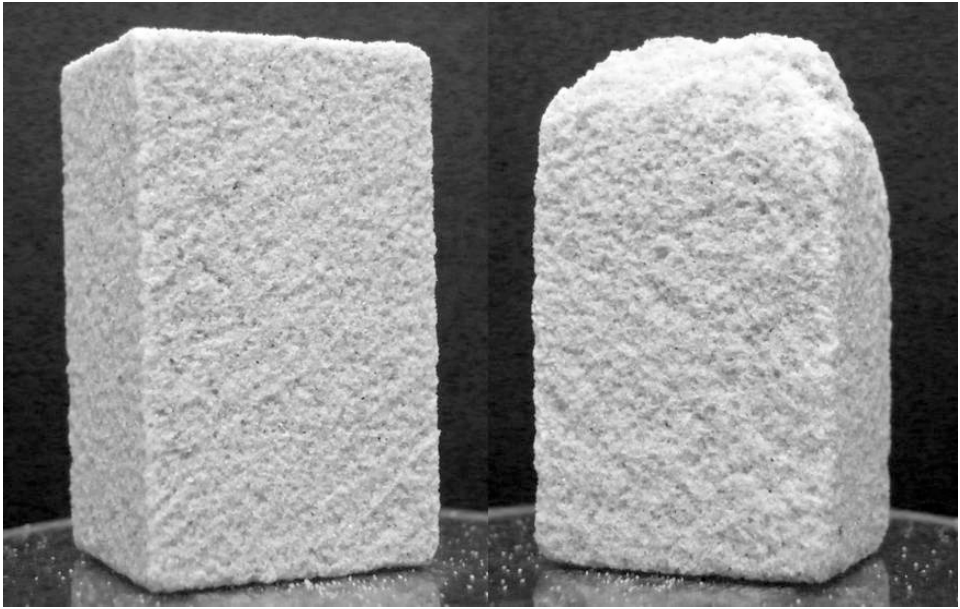

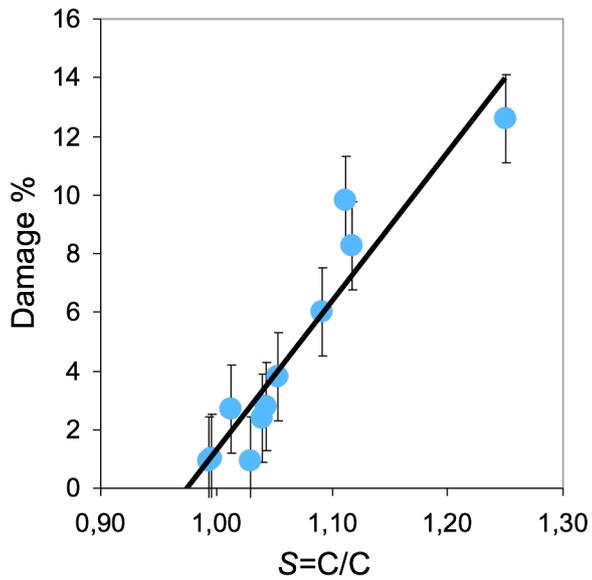

Figure 1 (a) Prague Sandstones before (on the left) and after (on the right) 2 wetting and drying cycles at 21°C (b) Damage $\left(m_{stone}/m_0\right) \times 100$ observed after the second cycle (rewetting with water/drying) as a function of the supersaturation of the salt solution with respect to mirabilite used in the first cycle (wetting with salt solution/drying).

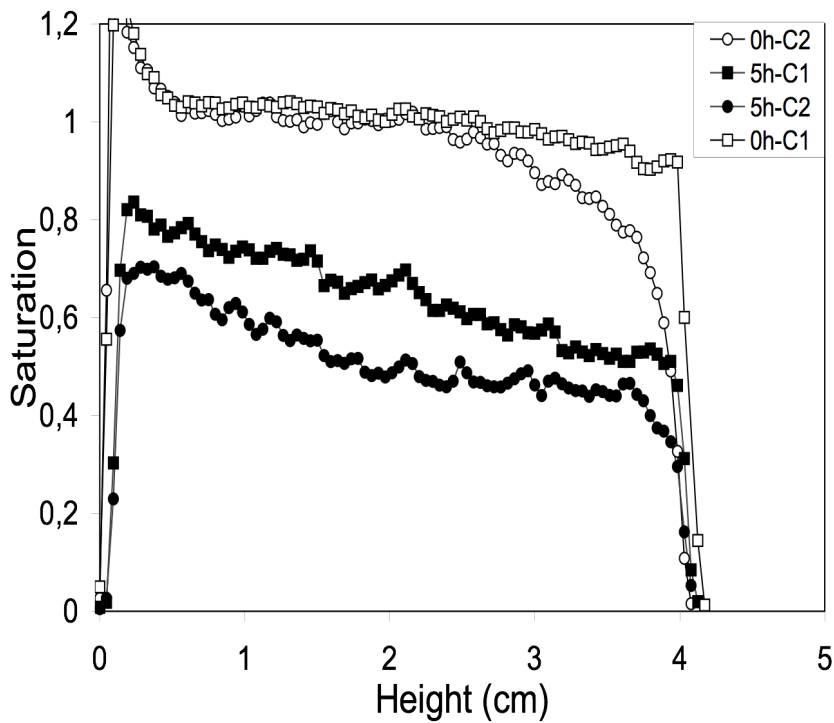

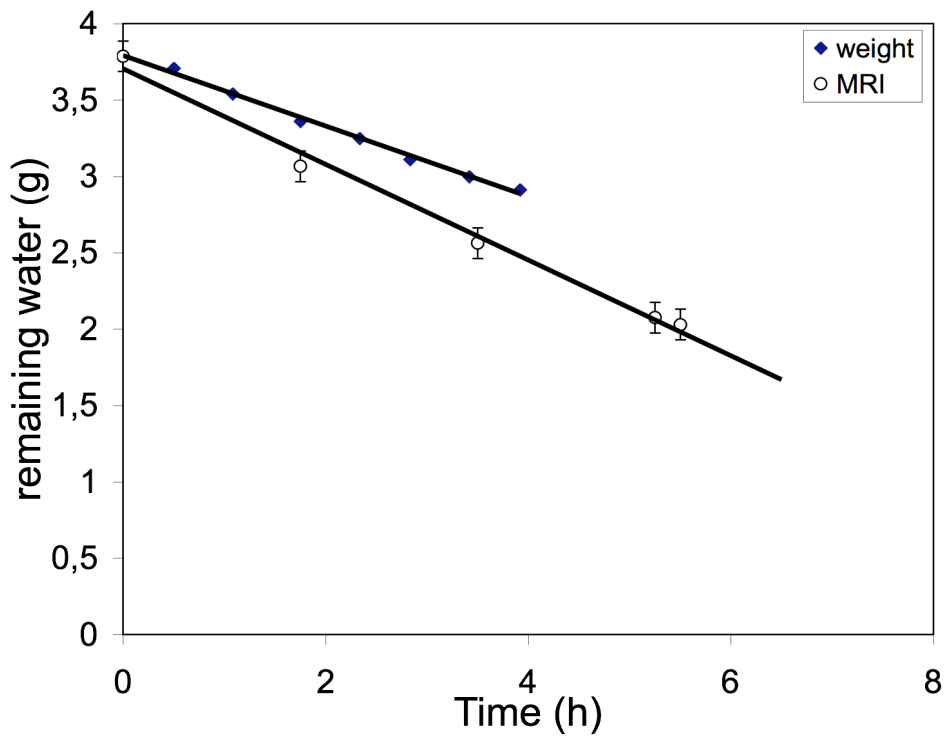

Figure 2a. MRI saturation profiles during drying; the sample was imbibed with sodium sulfate solution at S=1 for the first cycle. Squares are for cycle 1, circles for cycle 2; a comparison is made between the time instants just after imbibition and after 5 hours of drying. (b) Difference between water content from the MRI measurements (open symbols) and the weight measurements (filled symbols) during cycle 2. The difference is due to the formation of hydrated crystals that are not visible as water in the MRI experiments.

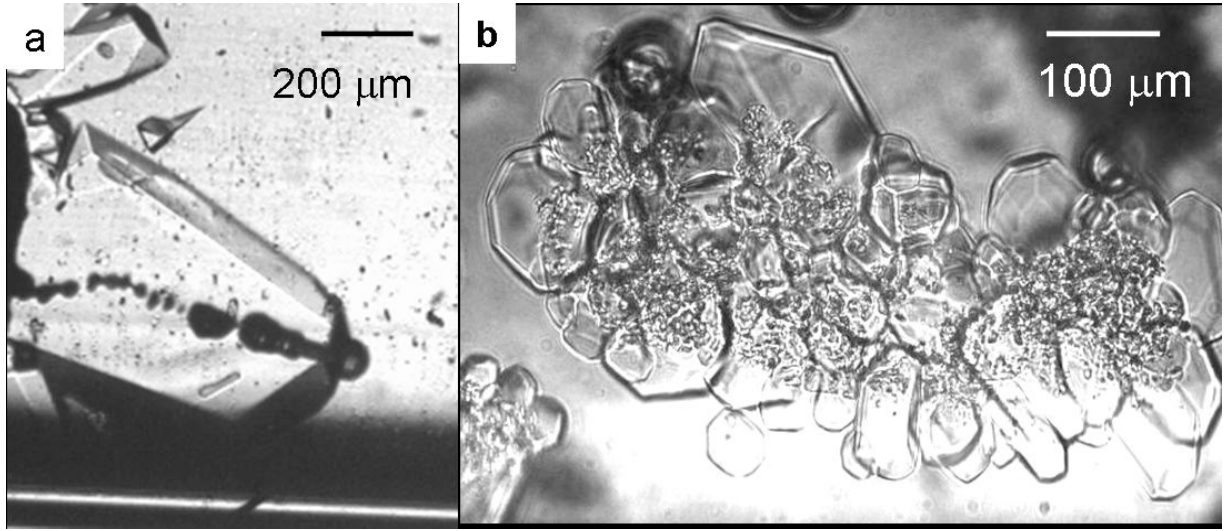

Figure 3. Microphotographs of the formation of hydrated crystals mirabilite) in a rectangular capillary. (a) Slow growth of large crystals from solution in cycle 1. (b) After rewetting with pure water (cycle 2), the thenardite crystallites that are visible as the small grains in the center of the aggregate do not dissolve completely, and act as nucleation centers for growth of the hydrated crystals. The latter are visible as the facetted transparent crystals.

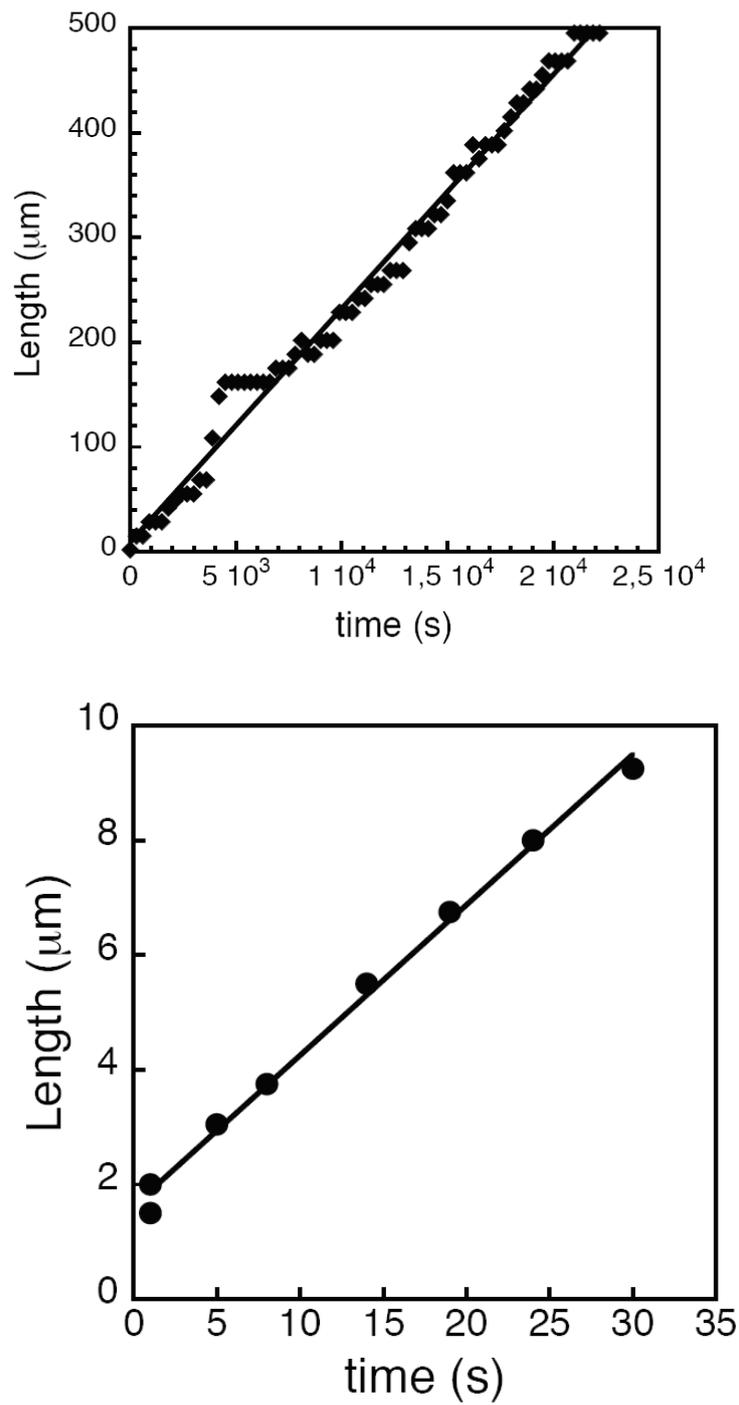

Fig.4 Dynamics of crystallization: typical size of a mirabilite crystal forming during the first cycle as a function of time. Inset: the much faster growth velocity when the mirabilite crystals form as observed in Fig.3.